\documentclass[twocolumn,a4paper,10pt,aps,prd,amsmath,amssymb,preprintnumbers,longbibliography]{revtex4-1}
\usepackage{amsmath} 
\usepackage{amsfonts} 
\usepackage{amssymb}
\usepackage{bbm}
\usepackage{graphics}
\usepackage{graphicx}
\usepackage{titlesec}
\usepackage{mathtools}
\usepackage{environ}
\usepackage{dsfont}

\usepackage[colorlinks=true,hyperindex,breaklinks]{hyperref}
\hypersetup{urlcolor=black,linkcolor=black,citecolor=black}
\usepackage[toc,page]{appendix}

\usepackage{makecell,tabularx}
\setcellgapes{3pt}

\makeatletter
\renewcommand*\env@matrix[1][c]{\hskip -\arraycolsep
  \let\@ifnextchar\new@ifnextchar
  \array{*\c@MaxMatrixCols #1}}
\makeatother

\newcommand{\be}{\begin{equation}}
\newcommand{\ee}{\end{equation}}
\newcommand{\ba}{\begin{eqnarray}}
\newcommand{\ea}{\end{eqnarray}}

\titleformat{\subsection}[block]{\normalfont\bfseries}{\thesubsection.}{1ex}{}
\titlespacing{\subsection}{0pt}{10pt}{1pt}[0pt]
\titleformat*{\section}{\large\bfseries}
\renewcommand{\thesubsection}{\arabic{subsection}}

\usepackage{natbib} 
\setcitestyle{sort&compress,cite}

\usepackage{braket}

\usepackage{todonotes}
\definecolor{refkey}{rgb}{0,0,1}
\definecolor{labelkey}{rgb}{0,1,0}

\renewcommand{\thesection}{\arabic{section}}
\renewcommand{\thesubsection}{\thesection.\arabic{subsection}}

\makeatletter
\renewcommand{\p@subsection}{}
\renewcommand{\p@subsubsection}{}
\makeatother

\usepackage{enumitem}

\begin{document}

\title[ ]{Crossing the Big Bang singularity}

\author{C. Wetterich}
\affiliation{Institut  f\"ur Theoretische Physik\\
Universit\"at Heidelberg\\
Philosophenweg 16, D-69120 Heidelberg}

\begin{abstract}
A simple model for a scalar field and gravity admits homogeneous isotropic cosmological solutions which cross the Big Bang singularity. In the scaling frame with field dependent effective Planck mass these solutions are regular. They become singular in the Einstein frame with fixed Planck mass. This field singularity arises since the field transformation of the metric to the Einstein frame is singular at the crossing point of a vanishing scalar field. No physical singularity is present for these solutions. Within general models with no more than two derivatives we find that the possibility of a crossing of the "Big Bang singularity" is not generic. It needs a tuning of model parameters. The present models are not a realistic description of the Universe since they fail to render acceptable inhomogeneities.
\end{abstract}

\maketitle

It is commonly believed that cosmological models of the Big Bang have an initial singularity where our observed Universe shrinks to a point \cite{ROB,LIKH,Penrose1965,Hawking1966,BKL,MIS}.
This includes models of inflation 
\cite{Starobinsky1980,Guth1981,Mukhanov1981,Linde1982,Albrecht1982,Shafi1983,Linde1983}, for which moving massive particles "see" a finite proper time to the singularity \cite{Borde2001,Mithani2012} (geodesic incompleteness). Speculations that this "Big Bang singularity" may be crossed during the evolution of the Universe have been made by many authors 
\cite{TOL,BRVA,GAVE,KOS,STU1,STTU,KOSST,NABV,LFTS,TOWE,BOJO,TUPS,ACS,BUK,LETU,ASH,GUPE,PEN,CEB,BRABE,LES,Bars2013,IST,LEVY,GKR,%
BRAP,Kamenshchik2016,BASSTTU}. 
In such models the Universe undergoes a "shrinking stage" where it approaches the singularity, and expands after crossing the singularity. Not much is known about the precise conditions for this to happen, mostly because of the absence of concrete diffeomorphism invariant field equations admitting crossing solutions which can be studied explicitly.

It has been noticed that for many models the "Big Bang singularity" is actually not a physical singularity, but rather due to a singular choice of fields
\cite{CWUWE,Wetterich2013,Wetterich2014,Wetterich2014a,CWGE,PFF%
}.
For such field singularities there exist other choices of the metric or other "frames" for which the solutions remain regular. In particular, a crossing of the Big Bang singularity can be described by regular solutions. First examples of this type of controlled "singularity crossing" can be found in Ref.~\cite{Wetterich2014a}, see also ref.~\citep{Kamenshchik2016},~\citep{CKK}.

The models~\cite{Wetterich2014a} that describe 
explicitly
the crossing of the Big Bang singularity
by regular solutions of the field equations
involve higher derivative terms. The field equations are rather complex and make systematic studies cumbersome. In this note we present a simple model of a scalar field coupled to gravity for which the effective action contains no more than two derivatives. For this model of variable gravity the dynamical Planck mass is given by a scalar field $\chi$. The coefficient of the scalar kinetic term is negative without inducing any instability. For a simple solution $\chi$ crosses the value zero. In the Einstein frame this corresponds to a crossing of the Big Bang singularity. Our model constitutes a very simple explicit example where the evolution of the Universe crosses the Big Bang singularity. Due to the presence of a regular frame all questions related to the properties of the Big Bang crossing can be studied
directly in a well defined framework.
In particular, we will address the question if the models admitting a crossing cosmology are generic. We also briefly address the question if the properties of the primordial fluctuation spectrum are realistic, and the issue of singularity for the neighboring anisotropic solutions. 

\section{Crossing Big Bang model}\label{sec:1}

\textit{Scaling frame.} Our model considers
a possible
quantum effective action for variable gravity
\cite{Wetterich2013}
\begin{equation}
\Gamma = \int_\chi \sqrt{g} \left\lbrace - \frac{\chi^2}{2} R + \frac{1}{2} (B-6) \partial^\mu \chi \partial_\mu\chi + \mu^4\right\rbrace.
\label{eq:1}
\end{equation}%
The Planck mass is dynamical, given by the scalar field $\chi$. The potential is simply a constant. For the kinetic term we take constant $B$. The model is stable for $B>0$ despite the fact that the coefficient of the kinetic term -- the kinetial -- is negative for $B<6$. This can be seen from the field equations below or by a transformation to the Einstein frame, also given below. The reason of stability is a mixing of the kinetic terms for $\chi$ and the scalar mode in the metric. We will choose $B=4$, or a negative kinetic term with $(B-6)/2=-1$. The negative kinetic term is important.

This model is a Brans-Dicke theory with a cosmological constant and its solutions are discussed in ref.~\cite{CWVN}, see also ref.~\cite{KPTVV}. The new point in this paper is not so much the particular solution presented in detail below, but rather the interpretation in terms of a Big Bang crossing in the associated Einstein frame. We do not propose here a realistic model for the Big Bang crossing, but rather a useful conceptual framework in which this issue can be investigated systematically.

We describe homogeneous isotropic cosmologies with a Robertson-Walker metric with scale factor $a(t)$, cosmic time $t$, Hubble parameter $H=\partial_t \ln a$, and curvature scalar $R = 12 H^2 + 6 \dot{H}$, with dots denoting the time derivatives, $\dot{H}=\partial_t H$. The gravitational field equations  obtain by variation of the effective action (\ref{eq:1}) with respect to the metric. They read
\cite{Wetterich2013}
\begin{equation}
3 \chi^2 H^2 + 6H \chi\dot{\chi} = \mu^4 + \frac{B-6}{2}\dot{\chi}^2,
\label{eq:2}
\end{equation}%
and
\begin{equation}
\chi^2 R = 4\mu^4 - B \dot{\chi}^2 - 6\chi(\ddot{\chi} + 3H\dot{\chi}).
\label{eq:3}
\end{equation}%
The scalar field equation is given by
\begin{equation}
(B-6)(\ddot{\chi} + 3H\dot{\chi}) = \chi R.
\label{eq:4}
\end{equation}%
For the particular choice $B=4$ the three field equations are solved for flat Minkowski space with constant scale factor $\bar{a}$,
\begin{equation}
g_{\mu\nu} = \bar{a}^2 \eta_{\mu\nu},\quad H=0, \quad R=0,\quad \dot{\chi} = \mu^2.
\label{eq:5}
\end{equation}%

The evolution of the scalar field is very simple,
\begin{equation}
\chi = \mu^2 t,
\label{eq:6}
\end{equation}%
where we choose $t$ such that $\chi(t=0)=0$. This solution is invariant under time reflection accompanied by a change of sign for $\chi$, $t\rightarrow -t$, $\chi \rightarrow -\chi$.

There is not much doubt that this simple solution is regular for all times. (One could modify the model in order to stop the increase of $|\chi|$ for $|t|\rightarrow \infty$.) We will show next that the crossing of zero by the scalar field corresponds to the Big Bang singularity in the Einstein frame. This singularity is induced by a singular choice of the metric variable.

\textit{Einstein frame.} The metric in the Einstein frame $g_{\mathrm{E}\,\mu\nu}$ obtains from the metric in the scaling frame $g_{\mu\nu}$ used in Eq.\ (\ref{eq:1}) by a Weyl transformation
\cite{HW,DIC}
\begin{equation}
g_{\mathrm{E}\,\mu\nu} = w^2 g_{\mu\nu},\quad w^2 = \frac{\chi^2}{M^2}.
\label{eq:7}
\end{equation}%
The Planck mass $M$ is introduced only by the definition of $g_{\mathrm{E}\,\mu\nu}$, rather than being a fundamental mass scale. The transformation is singular for $\chi = 0$. In the Einstein frame the effective action (\ref{eq:1}) reads
\begin{equation}
\Gamma = \int_x \sqrt{g_\mathrm{E}} \left\lbrace - \frac{M^2}{2} R_\mathrm{E} + \frac{BM^2}{2\chi^2} \partial^\mu \chi \partial_\mu\chi + \frac{M^4\mu^4}{\chi^4}\right\rbrace,
\label{eq:8}
\end{equation}%
where indices are now raised with $g_\mathrm{E}^{\mu\nu}$, and $R_\mathrm{E}$ is the curvature scalar in the Einstein frame. Defining a different normalization of the scalar field,
\begin{equation}
\varphi = M \ln \left(\frac{\chi^4}{\mu^4} \right),
\label{eq:9}
\end{equation}%
the effective action takes the form
\begin{equation}
\Gamma = \int_x \sqrt{g_\mathrm{E}} \left\lbrace - \frac{M^2}{2} R_\mathrm{E} + \frac{B}{32} \partial^\mu \varphi \partial_\mu\varphi + V_\mathrm{E}(\varphi)\right\rbrace,
\label{eq:10}
\end{equation}%
with exponential potential
\begin{equation}
V_\mathrm{E}(\varphi) = M^4 \exp (-\frac{\varphi}{M}).
\label{eq:11}
\end{equation}%
We could absorb a factor $B/16$ by a rescaling of $\varphi$ in order to bring the kinetic term to a standard form, but we will not do this here. We observe that positive and negative values of $\chi$ correspond to the same $\varphi$. The value $\chi = 0$ corresponds to the limit $\varphi \rightarrow -\infty$.

On the level of the quantum effective action, and the field equations and propagators derived from it, it has been argued that all models related by field transformations are indistinguishable by observation and therefore completely equivalent \cite{CWVN,Wetterich1988}. This property has been called "field relativity" \cite{CWUWE,Wetterich2014a}. For many observable quantities explicit transformations between different frames and frame-invariant formulations have been established
\cite{CWVN,DAM,FHU,FAR,FLA,CPS,DESA,CHYA,Wetterich2014,POVO,Wetterich2015,JKSW,JKMR,KPT,KAPI,BASSTTU,
NMS,DNSW}.
We discuss here the solutions of the field equations derived from the effective action (\ref{eq:10}),(\ref{eq:11}). They are equivalent to the solutions obtained from the effective action (\ref{eq:1}), in particular the crossing solution (\ref{eq:5}),(\ref{eq:6}). Conformal time $\eta$ is frame invariant. Correspondingly, $\chi(\eta)$ is the same in both frames.

For arbitrary $B(\varphi)$ the gravitational field equations derived from the effective action (\ref{eq:8}) are for a Robertson-Walker metric given by
\begin{equation}
3M^2 H_\mathrm{E}^2 = V_\mathrm{E} + \frac{B}{32} \dot{\varphi}^2,
\label{eq:12}
\end{equation}%
and
\begin{equation}
M^2R_\mathrm{E} = M^2 (12H_\mathrm{E}^2 + 6\dot{H}_\mathrm{E}) = 4V_\mathrm{E} - \frac{B}{16} \dot{\varphi}^2.
\label{eq:13}
\end{equation}%
Dots denote now derivatives with respect to cosmic time $t_\mathrm{E}$ in the Einstein frame. The scalar field equation obtains as
\begin{equation}
\ddot{\varphi} + 3H\dot{\varphi} + \frac{1}{2}\,\frac{\partial \ln B}{\partial\varphi}\,\dot{\varphi}^2 = -\frac{16}{B}\,\frac{\partial V_\mathrm{E}}{\partial\varphi} = \frac{16 V_\mathrm{E}}{BM}.
\label{eq:14}
\end{equation}%
Eqs.\ (\ref{eq:12}) and (\ref{eq:13}) can also be written in the form
\begin{equation}
\frac{B}{32} \dot{\varphi}^2 = - M^2 \dot{H}_\mathrm{E},\quad V_\mathrm{E} = M^2 (3H_\mathrm{E}^2 + \dot{H}_\mathrm{E}).
\label{eq:15}
\end{equation}%

For a solution \cite{Wetterich1988}  we make the ansatz
\begin{equation}
H_\mathrm{E} = \frac{\bar{\eta}}{t_\mathrm{E}},\quad V_\mathrm{E} = \frac{c^2M^2}{t_\mathrm{E}^2},
\label{eq:16}
\end{equation}%
leading to
\begin{equation}
\dot{\varphi} = \frac{2M}{t_\mathrm{E}},\quad \dot{H}_\mathrm{E} = -\frac{1}{\bar{\eta}} H_\mathrm{E}^2.
\label{eq:17}
\end{equation}%
Insertion into the field equation yields for constant $B$
\begin{equation}
\bar{\eta} = \frac{B}{8},\quad c^2 = \frac{B}{8}\left( \frac{3B}{8} - 1 \right).
\label{eq:18}
\end{equation}%
In particular, for
\begin{equation}
B=4,\quad \bar{\eta}=\frac{1}{2},\quad c^2 = \frac{1}{4}
\label{eq:19}
\end{equation}%
one has
\begin{equation}
H_\mathrm{E} = \frac{1}{2t_\mathrm{E}},\quad \dot{H}_\mathrm{E} = -\frac{1}{2t_\mathrm{E}^2},\quad R_\mathrm{E} = 0,\quad \varphi = 2M \ln (2Mt_\mathrm{E}).
\label{eq:20}
\end{equation}%
For $t_\mathrm{E} \rightarrow 0$ the Hubble parameter diverges and the scalar field $\varphi$ moves to $-\infty$. This associates $t_\mathrm{E} \rightarrow 0$ with the crossing of zero by $\chi$ in the scaling frame, $\chi(t_\mathrm{E} \rightarrow 0) = 0$.

The solution (\ref{eq:20}) corresponds precisely to the solution (\ref{eq:5}),(\ref{eq:6}) for $\chi > 0$ in the scaling frame. With Eq. (\ref{eq:9}) one has
\begin{equation}
\chi = \mu\sqrt{2Mt_\mathrm{E}}.
\label{eq:eq1}
\end{equation}%
We need the translation to conformal time $\eta$. (The similar symbols for conformal time and the constant $\bar{\eta}$ in the ansatz (\ref{eq:16}) are due to historical conventions and should not lead to confusion.) With
\begin{equation}
a_\mathrm{E} = c_\mathrm{E}\sqrt{2Mt_\mathrm{E}},\quad \frac{\mathrm{d}\eta}{\mathrm{d}t_\mathrm{E}} = \frac{1}{a_\mathrm{E}},
\label{eq:eq2}
\end{equation}%
one finds
\begin{equation}
\eta = \frac{1}{c_\mathrm{E}M}\sqrt{2Mt_\mathrm{E}},\quad \chi = c_\mathrm{E}M\mu\eta.
\label{eq:eq3}
\end{equation}%
In the scaling frame one has for Minkowski space $t=\bar{a}\eta$. Eqs.\,(\ref{eq:eq3}) and (\ref{eq:6}) indeed agree if we adjust the free dimensionless integration constants $c_\mathrm{E}$ and $\bar{a}$ such that $c_\mathrm{E}M=\bar{a}\mu$.

\textit{Crossing the singularity.} We next establish that the solution (\ref{eq:20}) becomes singular for $t_\mathrm{E}\rightarrow 0$. While for general constant $B$ the curvature scalar diverges,
\begin{equation}
R_\mathrm{E} = \frac{6\bar{\eta}(2\bar{\eta}-1)}{t_\mathrm{E}^2} = \frac{3B(B-4)}{16t_\mathrm{E}^2},
\label{eq:21}
\end{equation}%
it vanishes for the special value $B=4$. What diverges for $B=4$ and $t_\mathrm{E} \rightarrow 0$ is the squared Ricci tensor
\begin{equation}
R_{\mathrm{E}\,\mu\nu} R_{\mathrm{E}}^{\mu\nu} = \frac{1}{4}\,R_\mathrm{E}^2 + 3\dot{H}_\mathrm{E}^2 = \frac{3}{4t_\mathrm{E}^4}.
\label{eq:22}
\end{equation}%
Usually, such a behavior is called a singularity.
With scale factor $a_\mathrm{E}$ in the Einstein frame
\begin{equation}
a_\mathrm{E} = d_0 \sqrt{t_\mathrm{E}},\quad \sqrt{g_\mathrm{E}} = d_0^3 t_\mathrm{E}^{3/2},\quad d_0 = c_\mathrm{E}\sqrt{2M},
\label{eq:23}
\end{equation}%
also the combination
\begin{equation}
\sqrt{g_\mathrm{E}}R_\mathrm{E}^{\mu\nu} R_{\mathrm{E}\,\mu\nu} = \frac{3d_0^3}{4}\,t_\mathrm{E}^{-5/2}
\label{eq:24}
\end{equation}%
diverges. Since we have seen before in the scaling frame that no physical singularity is associated to $\chi=0$, we already know that this singularity in the Einstein frame cannot be a physical singularity. It is a pure "field singularity", arising from the singular choice of the metric field $g_{\mathrm{E}\,\mu\nu}$ for $\chi \rightarrow 0$.

So far we have concentrated on the solution where the Universe expands as $t_\mathrm{E}$ increases, moving away from the singularity. There exists another solution where the Universe shrinks for increasing $t_\mathrm{E}$, evolving into the singularity. Using negative $t_\mathrm{E}$ the shrinking solution (Big Crunch) is given by ($B=4$)
\begin{align}
\begin{split}
&H_\mathrm{E} = \frac{1}{2t_\mathrm{E}},\quad \dot{H}_\mathrm{E} = -\frac{1}{2t_\mathrm{E}^2},\quad a_\mathrm{E} = d_0 \sqrt{-t_\mathrm{E}},\quad R_\mathrm{E} = 0,\\
&\varphi = 2M \ln (-2Mt_\mathrm{E}),\quad \dot{\varphi} = \frac{2M}{t_\mathrm{E}},\quad V_\mathrm{E} = \frac{M^2}{4t_\mathrm{E}^2}.
\end{split}\label{eq:25}
\end{align}%
As $t_\mathrm{E}$ increases this solution reaches the singularity at $t_\mathrm{E} = 0$. In this limit $\varphi$ moves to $-\infty$.

Seen from the regular scaling frame the interpretation of the two solutions (\ref{eq:20}),(\ref{eq:25}) is simple. For the solution (\ref{eq:6}) one starts for negative $t$ with negative $\chi$, reaches $\chi=0$ at $t=0$, and has positive $\chi$ for positive $t$. The part with negative $t$ and negative $\chi$ corresponds to the "Big Crunch" where the Universe shrinks and $\varphi$ decreases to $-\infty$. At $t=0$ the singularity is reached. For positive $t$ one has the Big Bang with an expanding Universe with $\varphi$ increasing. This describes a crossing of the singularity, from a Big Crunch to a Big Bang. We recall that cosmic time $t$ in the scaling frame is not the same as cosmic time $t_\mathrm{E}$ in the Einstein frame -- they are related by the common conformal time $\eta$. We have chosen time conventions where $t=0$ corresponds to $t_\mathrm{E}=0$, and $t>0$ ($t<0$) is mapped to $t_\mathrm{E}>0$ ($t_\mathrm{E}<0$).

These simple findings demonstrate our central statement: For homogeneous isotropic cosmology the crossing of a singularity in the Einstein frame can be a completely regular solution in some other frame. It is a field singularity, arising from a singular choice of "field coordinates".

We notice that we may add higher derivative geometric invariants to the effective action (\ref{eq:1}), as the squared curvature scalar, $\sqrt{g}\,R^2$, or the squared Weyl tensor, $W=\sqrt{g}\, C_{\mu\nu\rho\sigma} C^{\mu\nu\rho\sigma}$. Those terms, or possible generalizations of the type $\sqrt{g}\,RCR$ or $\sqrt{g}\,C_{\mu\nu\rho\sigma}DC^{\mu\nu\rho\sigma}$, with $C$ and $D$ functions of covariant Laplacians, are expected to be present in any type of quantum gravity. They will dominate the behavior of the metric propagator for momenta much larger than $\chi$. Such invariants play, however, no role for the solution (\ref{eq:6}). The invariant $W$ does not influence the homogeneous isotropic field equations since the Robertson-Walker metric is conformally flat. Also $\sqrt{g}\,R^2$ does not affect the solution. Its contribution to the field equation is proportional to $R$, and therefore vanishes by virtue of Eq.\ (\ref{eq:5}).
In a more general setting the higher derivative invariants may be important for the crossing solutions~\cite{Wetterich2014a}.

\section{Parameter tuning}\label{sec:2}

Let us next ask if a crossing of the Big Bang is generic, or if it needs the selection of particular models. We consider families of neighboring models and ask if a particular tuning of parameters is necessary for the existence of solutions for which the "Big Bang singularity" is crossed.

\textit{Frame invariant field equations.} It is convenient to discuss this issue by using frame invariant field equations. This permits the straightforward translation of a given solution from one frame to the other. We consider here the general form of the effective action for variable gravity with up to two derivatives \cite{Wetterich2013}
\begin{equation}
\Gamma = \int_\chi \sqrt{g} \left\lbrace -\frac{F(\chi)}{2}\,R + \frac{1}{2}\,K(\chi)\partial^\mu \chi \partial_\mu \chi + V(\chi) \right\rbrace.
\label{eq:pt1}
\end{equation}%
The field equations can be written in a frame invariant form \cite{Wetterich2015}. With conformal time $\eta$ the two gravitational field equations read
\begin{align}
2 \hat{\mathcal{H}}^2 + \partial_\eta \hat{\mathcal{H}} &= A^2 \hat{V},\label{eq:pt3}\\
\hat{\mathcal{H}}^2 - \partial_\eta\hat{\mathcal{H}} &= \frac{\hat{K}}{2}\,(\partial_\eta\chi)^2,\label{eq:pt4}
\end{align}%
and the scalar field equation is given by
\begin{equation}
\hat{K}(\partial_\eta^2 + 2\hat{\mathcal{H}}\partial_\eta)\chi + \frac{1}{2}\,\frac{\partial \hat{K}}{\partial\chi} (\partial_\eta\chi)^2 + A^2\,\frac{\partial\hat{V}}{\partial\chi} = 0.
\label{eq:pt5}
\end{equation}%

Here the frame invariant quantities are
\begin{align}
A &= \sqrt{F}\,a,\quad \hat{\mathcal{H}} = \partial_\eta \ln A, \label{eq:pt6} \\
\hat{V} &= \frac{V}{F^2},\quad \hat{K} = \frac{K}{F} + \frac{3}{2F^2} \left( \frac{\partial F}{\partial\chi} \right)^2.\label{eq:pt7}
\end{align}%
In the scaling frame one has $F=\chi^2$ and therefore
\begin{align}
\begin{split}
A &= a\chi,\quad \hat{\mathcal{H}} = \partial_\eta \ln a + \partial_\eta \ln\chi,\\
\hat{V} &= \frac{V}{\chi^4},\quad \hat{K} = \frac{B}{\chi^2}.
\end{split}\label{eq:pt8}
\end{align}%
The Einstein frame is characterized by $F=M^2$. Solutions for $\chi(\eta)$ and $A(\eta)$ of the field equations (\ref{eq:pt3})--(\ref{eq:pt5}) are valid in all frames related by Weyl scalings.

Let us define
\begin{equation}
\hat{B} = \hat{K}\chi^2,
\label{eq:pt8a}
\end{equation}%
such that $\hat{B}=B$ in the scaling frame. Combining Eqs.\ (\ref{eq:pt3})(\ref{eq:pt4}) one has
\begin{equation}
3\hat{\mathcal{H}}^2 = A^2\hat{V} + \frac{\hat{B}}{2\chi^2}\,(\partial_\eta \chi)^2.
\label{eq:pt8b}
\end{equation}%
Multiplying Eq.\ (\ref{eq:pt5}) by $\chi^2$ and inserting Eq.\ (\ref{eq:pt8b}) yields
\begin{align}
\begin{split}
\hat{B} &\left\lbrace \partial_\eta^2 \chi \pm 2 \partial_\eta\chi \sqrt{\frac{A^2\hat{V}}{3} + \frac{\hat{B}}{6\chi^2}\,(\partial_\eta\chi)^2}\right. \\
&+ \left.\frac{1}{2\chi} \left( \frac{\partial\ln\hat{B}}{\partial\ln\chi} -2 \right)(\partial_\eta \chi)^2 \right\rbrace
= -A^2 \chi^2 \frac{\partial\hat{V}}{\partial\chi}.
\end{split}
\label{eq:pt8c}
\end{align}%

\textit{Crossing solutions.} We are interested in crossing solutions $\chi(\eta)$ with
\begin{equation}
\chi(0) = 0,\quad \partial_\eta \chi = \xi_{0},
\label{eq:ptic}
\end{equation}%
with finite $\xi_{0}>0$. 
The scalar field changes from negative values for $\eta < 0$ to positive values for $\eta > 0$. If for the scaling frame with $F=\chi^2$ the geometry corresponding to this solution is regular, it will typically become singular in the Einstein frame where $F=M^2$. A first condition for such a solution is its existence for small values of $|\chi|$. We assume here that $\hat{V}$ and $\hat{B}$ behave near $\chi=0$ as
\begin{equation}
\hat{V} = \left( \frac{\mu }{\chi} \right)^{\gamma_0},\quad \hat{B}(0) = B_0\;.
\label{eq:pt8d}
\end{equation}%
For given $B_{0}$ we will find consistent crossing solutions for
\begin{equation}
\gamma_{0}=1\pm\sqrt{1+2B_{0}}.
\label{40A}
\end{equation}
This amounts to a tuning of model parameters.

In particular, in the Einstein frame with a canonical scalar kinetic term one has $F=M^{2}$, $K=1$, and therefore $V=M^{4}\widehat{V}$, $\widehat{K}=M^{-2}$, $\widehat{B}=\chi^{2}/M^{2}$.
This implies $B_{0}=0$, such that consistent crossing solutions can be found for
\begin{equation}
V(\chi\to 0)=\begin{cases}
V_{0} & (\gamma_{0}=0)\\
M^{4}\mu^{2}/\chi^{2}& (\gamma_{0}=2)
\end{cases}.
\label{40B}
\end{equation}

For the behavior~\eqref{eq:pt8d} one finds that
$\hat{\mathcal{H}}$ diverges for $\eta \rightarrow 0$, $\chi \rightarrow 0$. This divergence has to be canceled by the other terms in Eq. (\ref{eq:pt8c}), requiring
\begin{equation}
B_0\xi_{0} \left\lbrace \xi_{0}\mp 2 \sqrt{\frac{A^2\chi^2\hat{V}}{3} + \frac{B_0 \xi_{0}^2}{6}}\,\right\rbrace = A^2\chi^2\,\frac{\partial\hat{V}}{\partial\ln\chi}.
\label{eq:pte}
\end{equation}%
Here we assume $B_0>0$, finite $\partial\hat{B}/\partial\chi$ and finite $\partial_\eta^2 \chi$. Eq. (\ref{eq:pte}) is understood as the limit $\chi \rightarrow 0$ and the plus (minus) sign applies to $\hat{\mathcal{H}}>0$ ($\hat{\mathcal{H}}<0$). 
Eq.\,(\ref{eq:pte}) will yield a first relation between the model parameters and some particular combination of field values at $\eta=0$. We will below derive a second condition for the geometry to be regular at the crossing point in an appropriate frame. The two conditions will over-constrain the system, such that only for a particular relation or "tuning" between the model parameters $B_0$ and $\gamma_0$ a regular crossing solution can be obtained.

We next exploit the constraint (\ref{eq:pte}).
Assuming finite
\begin{equation}
\gamma(\chi) = -\frac{\partial\ln\hat{V}}{\partial\ln\chi},\quad \gamma(\chi=0) = \gamma_0,
\label{eq:ptf}
\end{equation}%
one obtains for the variable $z = A^2\chi^2\hat{V}$ the limiting condition for $\chi \rightarrow 0$, $\eta \rightarrow 0$,
\begin{align}
\begin{split}
B_0 \xi_{0} \left\lbrace \xi_{0}\mp 2 \sqrt{\frac{z_0}{3} + \frac{B_0\xi_{0}^2}{6}}\,  \right\rbrace = -\gamma_0 z_0,\\
z = A^2\chi^2\hat{V},\quad z(\eta=0) = z_0.
\end{split}\label{eq:ptg}
\end{align}%
Close to $\chi = 0$ we can approximate $\hat{V}$ by Eq. (\ref{eq:pt8d}) such that the finiteness of $z$ requires
\begin{equation}
A^2 \sim \chi^{\gamma_0 -2},\quad \hat{\mathcal{H}} = \frac{(\gamma_0 -2)\xi_{0}}{2\eta}.
\label{eq:pti}
\end{equation}%
For any initial $\partial_\eta\chi(\eta=0)=\xi_{0}$ the relation (\ref{eq:ptg}) fixes $z_0$ and therefore the behavior of $A$ as
\begin{equation}
z_0 = \frac{2B_0^2\xi_{0}^2}{3\gamma_0^2} \left\lbrace 1 - \frac{3\gamma_0}{2B_0} \pm \sqrt{1 + \frac{3(\gamma_0^2 - 2\gamma_0)}{2B_0}} \right\rbrace.
\label{eq:ptj}
\end{equation}%
This expresses the combination $z_0/\xi_{0}^2$ as a function of the model parameters $\gamma_0$ and $B_0$, constituting the first constraint.
For $\hat{V}>0$, $\gamma_0>0$ one has the condition $z_0>0$. The expression in the brackets needs to be positive. From Eq. (\ref{eq:ptg}) we learn that this is only possible for
\begin{equation}
z_0 > \frac{3}{4} \left( 1 - \frac{2B_0}{3} \right) \xi_{0}^2.
\label{eq:ptl}
\end{equation}%
The limiting solution with $z_0=0$ exists only for $B_0=3/2$.

Let us consider first the particular models (\ref{eq:1}) with constant $B$, corresponding to $\gamma_0=\gamma=4$, $B_0=B$. With
\begin{equation}
\frac{z_0}{\xi_{0}^2} = \frac{B^2}{24} \left\lbrace 1 - \frac{6}{B}\, \pm \sqrt{1 + \frac{12}{B}} \right\rbrace
\label{eq:pto}
\end{equation}%
the only possible solution needs the plus sign for the square root if $B<6$. Positive $z_0$ requires
\begin{equation}
B > \frac{3}{2}.
\label{eq:ptq}
\end{equation}%
We conclude that a crossing of the Big Bang singularity is only possible for $B>3/2$. By an analogue discussion this statement extends to more general models with arbitrary $\gamma_0>0$, with $B$ replaced by $B_0$.

\textit{Tuning of parameters.} We next derive a second constraint for $z_0/\xi_{0}^2$. For general model parameters $\gamma_0$, $B_0$ this will overconstrain the system, thus allowing only for particular combinations.
For $B_0>3/2$ we can implement at $\eta=0$ the initial conditions (\ref{eq:ptic}) for $\chi(\eta)$. We need to discuss the behavior of the second variable for $\eta \rightarrow 0$. Instead of the variable $A$ which diverges we could use $z$, with initial condition $z(0)=z_0$ given by Eq. (\ref{eq:ptj}). We equivalently employ
\begin{align}
\begin{split}
h &= \chi\hat{\mathcal{H}} = \pm \sqrt{\frac{z}{3} + \frac{\hat{B}}{6}\,(\partial_\eta\chi)^2},\\
z &= 3h^2 - \frac{\hat{B}}{2}\,(\partial_\eta\chi)^2.
\end{split}
\label{eq:ptr}
\end{align}%
The two frame invariant functions characterizing the general homogeneous and isotropic cosmologies are therefore $h(\eta)$ and $\chi(\eta)$, where $h(\eta)$ encodes the behavior of geometry. In terms of $h$ we
write the evolution equation (\ref{eq:pt8c}) in the form
\begin{align}
\begin{split}
\partial_\eta^2 \chi = \frac{1}{\chi} &\left\lbrace \left( 1 - \frac{\partial\ln\hat{B}}{\partial\ln\chi} - \frac{\gamma(\chi)}{2} \right) (\partial_\eta\chi)^2 \right. \\
	&-2h \partial_\eta \chi + \left.\frac{3\gamma(\chi)h^2}{\hat{B}} \right\rbrace.
\end{split}
\label{eq:pts}
\end{align}%
Here $\gamma(\chi) = -\partial\ln\hat{V}/\partial\ln\chi$
and $\hat{B}(\chi)$ are
known finite functions for a given model. The initial condition,
\begin{equation}
h(0) = h_0 = \pm \sqrt{\frac{z_0}{3} + \frac{B_0 \xi_{0}^2}{6}},
\label{eq:ptu}
\end{equation}%
guarantees that the bracket
on the r.\,h.\,s.\ of Eq.\,(\ref{eq:pts})
 vanishes for $\chi \rightarrow 0$, such that $\partial_\eta^2 \chi$ is defined for $\eta=0$, $\chi=0$.

The evolution equation for $h$ reads
\begin{equation}
\partial_\eta h = \hat{\mathcal{H}}\partial_\eta\chi + \chi\partial_\eta\hat{\mathcal{H}} = \frac{1}{\chi} \left[ h^2 + h\partial_\eta\chi - \frac{\hat{B}}{2}(\partial_\eta\chi)^2 \right].
\label{eq:ptv}
\end{equation}%
For a regular crossing solution the r.\,h.\,s.\, of Eq.\ (\ref{eq:ptv}) needs to be well defined for $\chi\rightarrow 0$, which imposes for the initial conditions
\begin{equation}
h_0^2 + h_0\xi_{0} - \frac{B_0}{2}\xi_{0}^2 = 0.
\label{eq:ptw}
\end{equation}%
Combined with Eq.\ (\ref{eq:ptu}) one has to require
\begin{equation}
\frac{z_0}{3} \pm \xi_{0}\sqrt{\frac{z_0}{3} + \frac{B_0\xi_{0}^2}{6}} - \frac{B_0\xi_{0}^2}{3} = 0.
\label{eq:ptx}
\end{equation}%
Dividing by $\xi_{0}^2$ yields our second constraint for $z_0/\xi_{0}^2$.

With Eqs.\ (\ref{eq:ptg}) and (\ref{eq:ptx}) we have two equations for $z_0/\xi_{0}^2$. They can be obeyed simultaneously only if $B_0$ and $\gamma_0$ are subject to a certain condition. Eliminating the square root by use of Eq.\ (\ref{eq:ptg}) one finds
\begin{equation}
\frac{z_0}{\xi_{0}^2} = \left( B_0-\frac{3}{2}\right)\left( 1+\frac{3\gamma_0}{2B_0}\right)^{-1}.
\label{eq:pty}
\end{equation}%
Reinserting this into Eq.\ (\ref{eq:ptx}) yields two possible solutions
\begin{equation}
B_0 = \gamma_0 \left( \frac{\gamma_0}{2} -1\right),\quad \frac{z_0}{\xi_{0}^2} = \frac{(\gamma_0+1)(\gamma_0-2)(\gamma_0-3)}{2(\gamma_0-1},
\label{eq:ptz}
\end{equation}%
or
\begin{equation}
B_0 = \frac{3}{2},\quad z_0 = 0,\quad h_0= \pm\frac{\xi_{0}}{2}.
\label{eq:psa}
\end{equation}%
In particular, for $\gamma_0=4$ the constraint (\ref{eq:ptz}) yields $B_0=4$ and we recover the solution (\ref{eq:6}) for constant $\gamma$ and $B$. 
For $\xi_{0}=0$ there are additional possibilities
\begin{equation}
\xi_{0}=0,\quad z_0=0,\quad h_0=0,
\label{eq:psb}
\end{equation}%
that will not be discussed in detail here.

We conclude that the crossing of the "Big Bang singularity" at $\chi=0$ with a finite nonzero derivative $\partial_\eta\chi=\xi_{0}$ at the crossing point is only possible for a selected class of models. 
%
Crossing solutions with $z_0 \neq 0$ require a tuning of model parameters according to the constraint (\ref{eq:ptz}). 
We concentrate on this generic case. The main result of this section is the observation that for general families of models parametrized by $\gamma_0$ and $B_0$ only a one-dimensional subclass is compatible with regular crossing solutions. For this subclass $B_0$ needs to be tuned in accordance to the behavior of the potential encoded in $\gamma_0$,
\begin{equation}
B_0 = \frac{\gamma_0^2}{2} - \gamma_0.
\label{eq:55a}
\end{equation}%

For an analytic form of $F(\chi)$ and $V(\chi)$ at $\chi=0$ both $F$ and $V$ involve integer powers of $\chi$ in an expansion around $\chi=0$. In consequence, $\gamma_0$ is an integer. For models with a discrete symmetry $\chi \rightarrow -\chi$ this integer is even. As a result, $B_0$ has also to be an even integer. This illustrates in a simple way that crossing solutions need a specific choice of parameters. Inversion of eq.~\eqref{eq:55a} yields eq.~\eqref{40A}.

\textit{Generic models.} For generic models with a powerlike behavior of the potential near $\chi =0$, as encoded in $\gamma_0$, the relation (\ref{eq:55a}) is not met. If $B_0$ differs from the value (\ref{eq:55a}) and $B_0 \neq 3/2$, no solution can reach $\chi =0$ at finite $\eta_0$ with non-zero derivative $\partial_\eta\chi(\eta_0)$. One possibility is that $\chi =0$ is reached only for infinite $\eta \rightarrow \pm\infty$. This type of solution has been found for many models
\cite{Wetterich2013,Wetterich2014a,CWGE,PFF,Rubio2017}
that correspond to standard inflationary models in the Einstein frame. The physical picture is a beginning as the "great emptiness", rather than a crossing of the Big Bang singularity. As another possibility $\chi =0$ is not reached for any value of $\eta$. Solutions are repelled from the "Big Bang singularity". Finally, the approach of $\chi(\eta)$ to the value $\chi =0$ could be non-analytic. In this case one may ask if a regular behavior could be obtained by a different choice of the scalar field.

\section{Scaling solutions}\label{sec:3}

Models with constant $\gamma$ and $\hat{B}$ admit particularly simple scaling solutions. 
These scaling solutions permit for a more global view of candidate solutions crossing the Big Bang singularity. In the vicinity of $\chi =0$ the regular crossing solutions should obey the constraints developed above, with $\gamma_0 = \gamma$ and $B_0 = \hat{B}$ identified.

\textit{Dimensionless field equations.} For a discussion of the scaling solutions it is convenient to switch to dimensionless field equations. This form of the equations is also particularly appropriate for possible numerical solutions.
The field equations (\ref{eq:pt3})--(\ref{eq:pt5}) are valid for $\chi\neq 0$. We have omitted overall powers of $\chi$ multiplying these equations. For $\chi\neq 0$ we can use the dimensionless variable
\begin{equation}
s = \ln\left(\frac{\chi}{\mu}\right),\quad \partial_\eta s = \frac{\partial_\eta\chi}{\chi},\quad \partial_\eta^2 s = \frac{\partial_\eta^2\chi}{\chi} - (\partial_\eta s)^2.
\label{eq:pt9}
\end{equation}%
Eq.\ (\ref{eq:pt4}) becomes
\begin{equation}
\hat{\mathcal{H}}^2 - \partial_\eta \hat{\mathcal{H}} = \frac{\hat{B}}{2}(\partial_\eta s)^2,
\label{eq:pt10}
\end{equation}%
while eq.\ (\ref{eq:pt5}) takes the form
\begin{equation}
\hat{B}\partial_\eta^2 s + 2 \hat{\mathcal{H}}\partial_\eta s + \frac{1}{2}\,\frac{\partial\hat{B}}{\partial s}\,(\partial_\eta s)^2 + A^2\, \frac{\partial\hat{V}}{\partial s} = 0.
\label{eq:pt11}
\end{equation}%

The explicit factors of $A$ can be absorbed by replacing $\eta$ by a new variable $y$ obeying
\begin{equation}
\frac{\partial y}{\partial \eta} = A.
\label{eq:pt12}
\end{equation}%
With
\begin{align}
\begin{split}
\tilde{H} = \frac{\hat{\mathcal{H}}}{A} = \partial_y \ln A,\quad \partial_y \tilde{H} + \tilde{H}^2 = \frac{1}{A^2}\,\partial_\eta \hat{\mathcal{H}},\\
\partial_y s = \frac{1}{A}\,\partial_\eta s,\quad \frac{1}{A^2}\,\partial_\eta^2 s = \partial_y^2 s + \tilde{H}\partial_y s,
\end{split}\label{eq:pt13}
\end{align}%
the geometric field equations become
\begin{align}
\partial_y \tilde{H} + 3\tilde{H}^2 &= \hat{V},
\label{eq:pt14}\\
\partial_y \tilde{H} &= - \frac{\hat{B}}{2} (\partial_y s)^2,
\label{eq:pt15}
\end{align}%
while the scalar field equation reads
\begin{equation}
\hat{B} \left[ \partial_y^2 s + 3\tilde{H}\partial_y s  + \frac{1}{2}\,\frac{\partial\ln B}{\partial s}\,(\partial_y s)^2 \right] + \frac{\partial\hat{V}}{\partial s} = 0.
\label{eq:pt16}
\end{equation}%
The quantities $y$, $\tilde{H}$, $s$, $\hat{V}$ and $\hat{B}$ are all dimensionless. The system of field equations (\ref{eq:pt14})--(\ref{eq:pt16}) has eliminated all dependence on frame and all scales. The overall size of $A$ plays no role.

It is instructive to see the simple crossing solution (\ref{eq:6}) in this formulation. The corresponding model is defined by
\begin{equation}
\hat{V} = e^{-4s},\quad \frac{\partial\hat{V}}{\partial s} = -4\hat{V},
\label{eq:pt17}
\end{equation}%
and $\hat{B}=B=4$. For the solution (\ref{eq:6}) one has in the scaling frame $\eta=t$, $\bar{a}=1$ and
\begin{align}
\begin{split}
\chi &= \mu^2 \eta,\quad s= \ln (\mu\eta),\\
A &= \chi,\quad \frac{\partial y}{\partial \eta} = \mu^2\eta,\quad y = \frac{\mu^2\eta^2}{2}.
\end{split}\label{eq:pt18}
\end{align}%
The frame invariant quantities obey
\begin{equation}
s = \frac{1}{2}\,\ln(2y),\quad \partial_y s = \frac{1}{2y},\quad \hat{V} = \frac{1}{4y^2},
\label{eq:pt19}
\end{equation}%
and
\begin{equation}
\tilde{H} = \frac{1}{\chi}\,\partial_\eta\ln\chi = \frac{1}{A}\,\partial_\eta s = \partial_y s.
\label{eq:pt20}
\end{equation}%
For $B=4$ all equations (\ref{eq:pt14})--(\ref{eq:pt16}) are obeyed indeed.

\textit{General scaling solutions.}
We next turn to general constant $\hat{B}$. We also consider constant $\gamma$, as defined by
\begin{equation}
\hat{V} = e^{-\gamma s},\quad s= -\frac{1}{\gamma}\,\ln \hat{V}.
\label{eq:pt21}
\end{equation}%
For the scaling solution we make the ansatz
\begin{equation}
\hat{V} = \frac{c}{y^2},\quad \partial_y s = \frac{2}{\gamma y},\quad \tilde{H} = \frac{b}{y}.
\label{eq:pt22}
\end{equation}%
With this ansatz the field equations become algebraic equations for the constants $c$, $b$,
\begin{align}
\begin{split}
b &= \frac{2\hat{B}}{\gamma^2},\\
c &= -b + 3b^2 = - \frac{2\hat{B}}{\gamma^2} + \frac{12\hat{B}^2}{\gamma^4}.
\end{split}\label{eq:pt23}
\end{align}%
Only two of the field equations are independent. The equation corresponding to Eq.\ (\ref{eq:pt16}),
\begin{equation}
\hat{B} \left( -\frac{2}{\gamma} + \frac{6b}{\gamma} \right) -c\gamma = 0,
\label{eq:pt23a}
\end{equation}%
is obeyed automatically if the two equations (\ref{eq:pt23}) are obeyed. For positive $\hat{V}$ we require
\begin{equation}
c > 0,\quad b> \frac{1}{3},\quad \hat{B}>\frac{\gamma^2}{6}.
\label{eq:pt24}
\end{equation}%

We next translate the scaling solution (\ref{eq:pt22}) to conformal time. For $b$ and $c$ given by Eq.\ (\ref{eq:pt23}) one has
\begin{equation}
\frac{\partial\ln A}{\partial y} = \frac{b}{y},\quad A = A_0 y^b,
\label{eq:pt25}
\end{equation}%
with $A_0$ an integration constant. The relation between $y$ and $\eta$ obtains as
\begin{equation}
\frac{\partial y}{\partial\eta} = A_0 y^b,\quad y = \left(\frac{\eta}{\eta_0} \right)^{\frac{1}{1-b}},
\label{eq:pt26}
\end{equation}%
where $\eta_0$ is related to the integration constant $A_0$ by
\begin{equation}
\eta_0 = \frac{1}{A_0(1-b)}.
\label{eq:pt27}
\end{equation}%

We also turn back to the scalar field $\chi$,
\begin{equation}
\ln\left(\frac{\chi}{\mu}\right) = s = -\frac{1}{\gamma}\,\ln\left(\frac{c}{y^2}\right),\quad \frac{\chi}{\mu} = \left(\frac{y^2}{c}\right)^\frac{1}{\gamma}.
\label{eq:pt28}
\end{equation}%
For the scaling solution it evolves with conformal time as
\begin{equation}
\chi = c^{-\frac{1}{\gamma}} \mu \left(\frac{\eta}{\eta_0}\right)^\frac{2}{\gamma(1-b)}.
\label{eq:pt29}
\end{equation}%
Finally, $A(\eta)$ and $\hat{\mathcal{H}}(\eta)$ are given by
\begin{equation}
A = A_0 \left(\frac{\eta}{\eta_0}\right)^\frac{b}{1-b},\quad \hat{\mathcal{H}} = \frac{b}{1-b}\,\eta^{-1}.
\label{eq:pt30}
\end{equation}%

Eqs.\ (\ref{eq:pt29}), (\ref{eq:pt30}) are the scaling solution in a frame invariant form. Eq.\ (\ref{eq:pt30}) can be translated into every particular frame by insertion of the definition of $A$. This yields the time dependence of the scale factor $a(\eta)$. For the example of the scaling frame one has
\begin{equation}
a = \frac{A}{\chi} = \bar{a}\left(\frac{\eta}{\eta_0}\right)^\frac{\gamma b - 2}{\gamma(1-b)},\quad \bar{a} = \frac{A_0 c^\frac{1}{\gamma}}{\mu}.
\label{eq:pt30a}
\end{equation}
In the Einstein frame with canonical scalar kinetic term the scaling solutions obey
\begin{equation}
a=\frac{A_{0}}{M}\left(\frac{\eta}{\eta_{0}}\right)^{\frac{b}{1-b}}\;,\quad \mathcal{H}=\partial_{\eta}\ln a=\frac{b}{1-b}|\eta |^{-1}.
\label{82A}
\end{equation}

The critical quantity for the behavior of the scaling solutions is
\begin{equation}
\delta = \frac{2}{\gamma(1-b)},\quad \chi\sim\eta^\delta.
\label{eq:pt31}
\end{equation}%
For $\delta < 0$ one finds that $\chi$ diverges for $\eta\rightarrow 0$ and vanishes for $\eta\rightarrow\infty$. 
A similar scaling solution exists for negative $\eta$, with $\chi =0$ reached only for $\eta\rightarrow -\infty$. For this type of solution the Universe is eternal -- it can be followed backwards to the infinite past. This is the type of solutions discussed in Refs.\,\cite{Wetterich2013,Wetterich2014a,CWGE,PFF}.
For $\delta < 0$
there is no crossing of the "Big Bang singularity" at $\chi=0$ for any finite $\eta$. 

For $\delta > 0$, however, $\chi=0$ is reached at $\eta=0$. The function $\chi(\eta)$ remains differentiable for integer $\delta$. For our example (\ref{eq:6}) $\gamma=4$, $B=4$, $b=1/2$ one has indeed $\delta=1$. Similarly, with
\begin{equation}
A \sim \eta^{\frac{\delta\gamma}{2}-1},
\label{eq:pt32}
\end{equation}%
one finds that $A$ vanishes for $\eta\rightarrow 0$ in a differentiable way for integer $\delta\gamma/2$ larger than one. A potential that is a polynomial in $\chi^2$ or $\chi^{-2}$ has even integer $\gamma$. For example, a potential $\sim\chi^2$ in the scaling frame has $\gamma = 2$. For $\delta\gamma = 2$ one has constant $A$, while $\delta\gamma > 2$ corresponds to $A(\eta=0)=0$. For the Einstein frame with canonical scalar kinetic term the crossing solution~\eqref{40B} with $\gamma_{0}=2$ has $b=0$, $\delta=1$, implying a constant scale factor $a$. With $\widehat{V}=\mu^{2}/\chi^{2}$, $B=\chi^{2}/M^{2}$ this matches the above scaling solution in the scaling frame with $\delta=1$.

\textit{Regular crossing scaling solutions.} The scaling solutions describing a regular crossing of the Big Bang singularity correspond to positive integer $\delta$.
Expressed in terms of the model parameters $\hat{B}$ and $\gamma$ one finds
\begin{equation}
\delta = \frac{2\gamma}{\gamma^2 - 2\hat{B}}.
\label{eq:pt33}
\end{equation}%
For $\delta =1$ this fixes $\hat{B}_1(\gamma)$,
\begin{equation}
\hat{B}_1(\gamma) = \gamma\left( \frac{\gamma}{2}-1\right).
\end{equation}%
We recognize the constraint (\ref{eq:ptz}). Since $B_1$ vanishes for $\gamma=0$ or $\gamma=2$, which corresponds in the scaling frame to a potential $\sim\chi^4$ or $\chi^2$, these cases are limiting cases.

Models with $\delta=2$ have to obey
\begin{equation}
\hat{B}_2(\gamma) = \frac{\gamma}{2}\,(\gamma-1).
\label{eq:pt35}
\end{equation}%
One has $\hat{B}_2(2)=1$, $\hat{B}_2(4)=6$. For a constant potential, $\gamma=4$, we observe two particular values of $\hat{B}$. For $\hat{B}=4$ one has $\chi\sim\eta$, $A\sim\eta$ and for $\hat{B}=6$ we find $\chi\sim\eta^2$, $A\sim\eta^3$. We conclude that the ansatz (\ref{eq:pt22}) leads to solutions that remain analytic for $\eta\rightarrow 0$ only for special values of $\hat{B}$.

\textit{General solutions in dimensionless variables.}
The ansatz (\ref{eq:pt22}) does not account for the most general solution of the system of differential equations (\ref{eq:pt14})--(\ref{eq:pt16}). The general solution $s(y)$, $\tilde{H}(y)$ has two free integration constants that we may take as $s(y_\mathrm{in})$, $\partial_\eta s(y_\mathrm{in})$. Indeed, we can combine the field equations into a simple equation for $s(y)$, with primes denoting derivatives with respect to $y$
\begin{equation}
\hat{B} \left( s'' \pm s'\sqrt{3\hat{V} + \frac{3\hat{B}}{2} s'^2} + \frac{1}{2}\,\frac{\partial\ln\hat{B}}{\partial s}\, s'^2  \right) = -\frac{\partial\hat{V}}{\partial s}.
\label{eq:pt36}
\end{equation}%
(Here the plus (minus) sign corresponds to positive (negative) $\tilde{H}$.) The function $\tilde{H}(y)$ can then be inferred from Eqs.\ (\ref{eq:pt14}), (\ref{eq:pt15}). One integration constant, say $s(y_\mathrm{in})$ can be absorbed by a shift in $y$. 
The general (local) solutions have therefore one additional free integration constant besides this trivial shift.

The scaling solutions are not the most general solutions. Indeed, the
ansatz (\ref{eq:pt25}) has no further integration constant and can therefore not be the most general solution. 
One may ask if for model parameters $\gamma$ and $\hat{B}$ not obeying the constraint for the crossing scaling solutions some of the general solutions could, nevertheless, cross the Big Bang singularity. This is not possible, however. The
constraints (\ref{eq:ptz})--(\ref{eq:psb}) tell us that generic solutions will not lead to a crossing of the "Big Bang singularity" with a finite derivative $\partial_\eta \chi$ at $\eta=0$. Solutions for which $\partial_\eta\chi$ remains always finite cannot cross the "Big Bang singularity" at $\chi=0$ unless one of the conditions (\ref{eq:ptz})--(\ref{eq:psb}) is obeyed.

\textit{Non-polynomial potentials.} Restrictions
on the allowed $\hat{B}$ for scaling solutions crossing the "Big Bang singularity" with finite $\partial_\eta\chi$ are connected to the assumption of integer values of $\gamma$ for a polynomial $\hat{V}(\chi)$. 
One may relax this assumption and also discuss non-polynomial potentials.
For a given $\hat{B}$ we can find a crossing scaling solution with $\delta=1$ provided $\gamma$ obeys the relation
\begin{equation}
\gamma = 1 \pm \sqrt{1+2B}.
\label{eq:psc}
\end{equation}%
In this case one has
\begin{equation}
\chi\sim\eta,\quad A\sim\eta^{\frac{\gamma}{2}-1}.
\label{eq:psd}
\end{equation}%
A crossing for a nonzero finite scale factor $a(0)=\bar{a}$ requires
\begin{align}
\begin{split}
F&\sim\chi^{\gamma-2}\sim\eta^{\gamma-2},\\
V &= \hat{V}F^2 = \chi^{\gamma-4}.
\end{split}\label{eq:pse}
\end{align}%

Allowing for non-polynomial $F$ and $V$ a scaling solution crossing $\chi=0$ with finite $\partial_\eta\chi$ is possible
for a continuous range in $\hat{B}$. A non-analytic form of $F$ and $V$ for $\chi\rightarrow 0$ may, however, not be the outcome of a quantum field theory. Even 
for non-polynomial $F$ and $V$
the crossing solutions are not generic. They still require a tuning between the parameters $\gamma$ and $\hat{B}$. 

A singular solution with $\chi\sim\eta^\delta$, and non-integer $\delta$ can be made regular by switching to a different scalar field $\sigma\sim\chi^{1/\delta}$. In this case a polynomial form of $V(\chi)$ and $F(\chi)$ transform to a non-analytic behavior of $V(\sigma)$ and $F(\sigma)$. 
The concept of regular solutions crossing the Big Bang employed in this note requires that both the functions $V(\chi)$ and $F(\chi)$ defining the model are regular at $\chi =0$, and the solution $\chi(\eta)$ is regular at $\chi =0$. With this understanding the
overall outcome of this investigation shows that crossing solutions are not generic for the class of variable gravity models (\ref{eq:pt1}). Tuning of model parameters is required.

Our frame invariant discussion and explicit solution allow an immediate translation to arbitrary metric frames. Among them are flat frames for which $R=0$, in particular the primordial flat frame~\cite{PFF} for which geometry becomes Minkowski space in the limit $\eta\to -\infty$. The conditions for $F$, $K$ and $V$ for which a primordial flat frame exists can be found in ref.~\citep{PFF}.

\section{Primordial fluctuations}\label{sec:4}

A reliable computation of the spectrum of primordial fluctuations for cosmologies crossing the Big Bang has been an open issue for a long time, giving rise to many debates. In the presence of an explicit model and explicit regular solutions of the field equations this spectrum can be computed in a very direct way. This assumes "Bunch-Davies-type" initial conditions, namely that for momenta much larger than the scale of the curvature or other geometric quantities the propagator approaches the vacuum propagator in Minkowski space.

The spectrum of of primordial fluctuations is given by the correlation function for the gauge invariant physical scalar and the graviton (traceless transverse tensor mode.) In turn, the correlation function is given by the inverse of the second functional derivative of the quantum effective action \cite{CWQCM,CWCF}. These simple relations permit the study of the cosmic fluctuations in a frame invariant way \cite{Wetterich2015}.

In particular, we can compute the fluctuation spectrum in the Einstein frame. The general model of variable gravity (\ref{eq:pt1}) with $F=\chi^2$ is given in the Einstein frame by Eq.\ (\ref{eq:10}), where $B$ is obtained as
\begin{equation}
B(\varphi) = \biggl( K (\chi(\varphi)) +6 \biggr) \left( 1- \frac{1}{4}\,\frac{\partial\ln V}{\partial\ln\chi}\right)^{-2}.
\label{eq:f1}
\end{equation}%
For this convention of the scalar field with an exponential potential (\ref{eq:11}) the properties of the fluctuation spectrum are directly related to the kinetial $B(\varphi)$. The tensor to scalar ratio $r$ and the scalar spectral index $n$ obtain as \cite{Wetterich2013,Wetterich2014a,Wetterich2019}
\begin{equation}
r=16\varepsilon,\quad n= 1-2\left(\varepsilon + M\,\frac{\partial\varepsilon}{\partial\varphi}\right),
\label{eq:f2}
\end{equation}%
where
\begin{equation}
\varepsilon = \frac{8}{B},
\label{eq:f3}
\end{equation}%
and $B$ has to be taken at the value of $\varphi$ where a given fluctuation crosses the horizon. 

A small $r$ requires rather large $B$. Our simple model with $B=4$ is not compatible with the small observed upper bound on $r$. It is actually not a model of slow roll inflation in the Einstein frame either. For an inflationary model the "slow roll parameter" $\varepsilon$ has to be small, in contrast to $\varepsilon=2$ for the simple crossing model.

The assumption of Bunch-Davies initial conditions is not guaranteed in a model crossing the Big Bang. Anisotropic solutions in the neighborhood of our solutions (and even some other neighboring homogeneous isotropic solutions) develop a singularity at the crossing $\chi\rightarrow 0$~\cite{LIKH,BKL,MIS,STAAN,FISA,KPVSTV}. For the model~\eqref{eq:1} this is due to a divergent graviton propagator in the limit $\chi\rightarrow 0$~\cite{CWGE,PFF}. It is possible that this disease can be cured by the inclusion of (non-polynomial) higher derivative terms in the effective action that lead to well behaved graviton and scalar propagators. Indeed, it is to be expected that the effective action cannot be approximated by the simple form~\eqref{eq:1} in field regions where propagators are not well behaved. As we have seen, higher derivative terms do not necessarily change the homogeneous solution. 

Another issue concerns the question if the anisotropic singularities are physical singularities or field singularities. For the relative fluctuations the singular behavior of the traceless transversal tensor fluctuations (gravitons) can be observed in a formulation that is invariant under conformal field transformations~\cite{Wetterich2015}.
If the relative anisotropic singularities were field singularities they could only be removed by disformal field transformations~\cite{DNSW,CKK}. 

We emphasize, however, that the frame invariant singularities concern the relative fluctuations in the linear approximation. They could diverge due to a correlation function remaining finite, while the scale factor vanishes~\cite{PFF}. In this case the linear approximation for relative fluctuations of the metric breaks down and one has to look for a new non-linear frame-invariant characterization of the metric fluctuations.
We do not further investigate these questions here since the present model is anyhow not necessarily a candidate for a realistic Universe.

\section{Discussion}\label{sec:5}

We have presented a simple model of variable gravity that admits a solution where the scalar field $\chi$ crosses the value $\chi=0$ at some conformal time $\eta=0$, $\chi(\eta=0)=0$, and with a finite derivative, $\partial_\eta\chi(\eta=0) = \xi_{0}$. Per se, this seems not to be very spectacular. Translated to the Einstein frame with fixed Planck mass this solution has a shrinking stage where the "Big Bang singularity" is approached, and subsequently an expanding stage where cosmology moves away from the singularity. The "Big Bang singularity" corresponds to the value $\chi=0$. Crossing this value corresponds to a crossing of the "Big Bang singularity" in the Einstein frame.

This finding clearly demonstrates that for this solution the character of the "Big Bang singularity" in the Einstein frame as a field singularity. Our solution is completely regular in the regular scaling frame of variable gravity. The singularity appears only by a field transformation to the Einstein frame. The corresponding Weyl transformation becomes singular at $\chi=0$. The Einstein frame is therefore a singular choice of field coordinates. The "Big Bang singularity" has a status comparable to a coordinate singularity in geometry. Only the coordinates are coordinates in the field space rather than coordinates for spacetime.

The simplicity of models of variable gravity with only two derivatives permits us to investigate the question if similar crossing solutions exist generically. Within this rather large class of models the answer is negative. Crossing solutions occur only in the case of particularly tuned parameters. For generic parameters the general cosmological solutions avoid a crossing of the "Big Bang singularity." More generic solutions reach the value $\chi=0$ only in the infinite past for $\eta \rightarrow -\infty$
\cite{Wetterich2013,Wetterich2014a,Rubio2017}, and may correspond to a crossover from an ultraviolet fixed point of quantum gravity for the infinite past to an infrared fixed point in the infinite future.

The present model and solution may not be a realistic description of a Universe crossing the Big Bang. As it stands, it fails to account for realistic anisotropies. This holds even if one assumes that somehow after the Big Bang the dynamical evolution of the correlation functions approaches Bunch-Davies initials conditions. 
If the crossing solution describes the inflationary epoch, the spectrum of primordial fluctuations conflicts with observation. Nevertheless, one can extend the model such that the inflationary epoch at finite $\chi$ follows a behavior different from the crossing for $\chi\to 0$. In this case more realistic models may be taylored.

For the model~\eqref{eq:1} realistic relative anisotropies diverge when extrapolated backwards to the Big Bang. This feature is due to the properties of the graviton propagator for $\chi\rightarrow 0$. It seems conceivable that a more realistic graviton propagator for $\chi=0$ can be realized in the presence of (non-polynomial) higher derivative terms in the effective action, curing thereby the divergence of the relative anisotropic solutions in the neighborhood of the isotropic crossing solution, which is found in the linear approximation.

It is also conceivable that crossing solutions could become more generic once higher derivative terms are included in the effective action. In this case one could perhaps also find crossing solutions that become realistic models of inflation in the Einstein frame. At the present stage there are many arguments disfavoring our Universe to be described by a crossing solution. Solutions where the Big Bang singularity in the Einstein frame corresponds to a regular fixed point in the infinite past in physical time seem more natural. Nevertheless, crossing cosmologies could remain a possibility, if higher derivative invariants play an important role for the crossing.

 






\bibliography{refs}

\end{document}